\documentclass[11pt]{article}
\usepackage{epsfig} 
\newcommand{\sect}[1]{ \section{#1} \setcounter{equation}{0} } 

\newcommand{\half}{\mbox{\small{$\frac{1}{2}$}}}

\newcommand{\threehalves}{\mbox{\small{$\frac{3}{2}$}}} 
\newcommand{\fivehalves}{\mbox{\small{$\frac{5}{2}$}}} 
\newcommand{\MSbar}{\overline{\mbox{MS}}} 
 
\newcommand{\Nc}{N_{\!c}}

\begin{document}

\title{Asymptotic freedom from the two loop term of the $\beta$-function in a
cubic theory}
\author{J.A. Gracey, \\ Theoretical Physics Division, \\ 
Department of Mathematical Sciences, \\ University of Liverpool, \\ P.O. Box 
147, \\ Liverpool, \\ L69 3BX, \\ United Kingdom.} 
\date{}

\maketitle 

\vspace{5cm} 
\noindent 
{\bf Abstract.} We renormalize a six dimensional cubic theory to four loops in 
the $\MSbar$ scheme where the scalar is in a bi-adjoint representation. The
underlying model was originally derived in a problem relating to gravity being 
a double copy of Yang-Mills theory. As a field theory in its own right we find
that it has a curious property in that while unexpectedly there is no one loop 
contribution to the $\beta$-function the two loop coefficient is negative. It 
therefore represents an example where asymptotic freedom is determined by the 
{\em two} loop term of the $\beta$-function. We also examine a multi-adjoint 
cubic theory in order to see whether this is a more universal property of these
models. 

\vspace{-18cm}
\hspace{13.25cm}
{\bf LTH 1233}

\newpage 

\sect{Introduction.}

Scalar $\phi^3$ theory in six dimensions has proved to be a useful laboratory
or tool to explore major ideas in quantum field theory. For instance, after the
discovery of asymptotic freedom in Quantum Chromodynamics (QCD), \cite{1,2}, it
was used as a testing ground to study the implications of this property. This
was because it was shown that six dimensional $\phi^3$ theory was 
renormalizable and also asymptotically free, \cite{3}. In other words the 
consequences of this characteristic of non-abelian gauge theories could be
explored in a simple environment without the complications of the gauge
structure. One reason why higher dimensional theories might be of interest for
lower dimensional ones is that the ultraviolet behaviour of one theory could be
related to the infrared dynamics of another, \cite{4,5}. That idea has yet to 
be realized concretely in the gauge theory context. However in the case of 
scalar $\phi^3$ theory, where the calculations were easier to carry out, it has
been used to explore certain infrared ideas associated with the strong 
interactions. For instance, it has proved useful as a toy model of Regge theory
but in six dimensions where ladder diagrams were analysed in order to gain 
insight into the Regge slope. Several articles in this direction in $\phi^3$ 
theory are \cite{6,7,8}, for example. Clearly there are limitations to such 
exercises. The obvious one is that of an unphysical spacetime dimension. A more
serious one is the lack of a bounded Hamiltonian which means that true bound 
state analyses could not be fully credible. Despite this the theory played a 
valuable role as a sounding board for exploring new ideas in a gauge theory. 
Other applications of cubic scalar theory lie mainly in condensed matter 
physics and in particular critical phenomena. For instance various decorations 
of the scalar field with different symmetries allowed the critical exponents 
that relate to percolation and the Lee-Yang edge singularity problems, 
\cite{9,10,11}, to be determined very accurately in the $\epsilon$ expansion in
integer dimensions below six.

More recently another perhaps surprising example of the connection a cubic 
scalar theory has with physics has emerged. In \cite{12,13} the idea that 
on-shell gravity could be interpreted as a double copy of Yang-Mills theory was 
initially put forward and generated a large amount of interest. For instance it
was shown that there was a relation between the product of Yang-Mills $n$-point
functions and the corresponding on-shell gravity ampltitudes representing a 
connection with the KLT relations, \cite{14}. This double copy of Yang-Mills 
appears to be widely accepted as an interesting interpretation and clearly is a
direction to pursue in the quest for a theory of quantum gravity. One 
consequence of the double copy vision was the connection with a scalar field 
endowed with a bi-adjoint symmetry, \cite{15,16}, although the gravity 
connection with a scalar cubic interaction was observed earlier in \cite{17}. 
For instance in \cite{15} it was shown that scattering amplitudes of the double
copy theory could be related to the gluonic ones in pure Yang-Mills. Another 
direction that was followed in \cite{18,19} was to study classical solutions of
a linearized version of Yang-Mills theory and their relation to double copies 
of scalar fields in the bi-adjoint cubic theory. These ideas were explored 
further in \cite{20,21} where new solutions were found further strengthening 
the double copy correspondence concept. While most of these studies were 
classical it is worth investigating the underlying quantum field theory in its 
own right to ascertain whether it has any other interesting properties. That is
the purpose of this article. In particular we will renormalize the theory to 
four loops in six dimensions and deduce the renormalization group functions. 
Doing so will reveal a curious feature. It will transpire that for general Lie 
groups the one loop coefficient of the $\beta$-function vanishes despite there 
being a one loop contribution to the $3$-point vertex function. This is a 
rather unusual and rare situation in a renormalizable theory although it is 
known that the one loop term of renormalization group functions, other than the
$\beta$-function, can be zero in other models. What makes this bi-adjoint 
theory even more intriguing is that the non-zero two loop $\beta$-function 
coefficient is negative. Therefore this model appears to be one of the few 
cases where the property of asymptotic freedom is determined purely from the 
{\em two} loop term of the $\beta$-function. We will explore some of the basic 
consequences of this property as well as finding the underlying reason why it 
emerges. As part of this investigation we will renormalize a generalization of 
the bi-adjoint model by allowing the field to take values in the adjoint 
representation of four (different) Lie groups which we will refer to as the 
quartic adjoint model.

The article is organized as follows. The background to the properties of the
bi-adjoint cubic scalar theory such as the group theory connected with the
computation are discussed in Section $2$. That will also include the details of
how we performed the computation the results of which are discussed in 
Section $3$. The generalization to the quartic adjoint model is provided in the
next section before the concluding remarks of Section $5$. An appendix records 
full details of the renormalization group functions of the quartic adjoint 
model. 

\sect{Background.}

To begin with we define the six dimensional Lagrangian for the bi-adjoint
cubic scalar theory that was derived from solutions of linearized Yang-Mills
theory and related to the double copy of gravity. If we denote the basic scalar
field by $\phi^{a_1 a_2}$ then the Lagrangian is, for example from 
\cite{19,20},
\begin{equation}
L ~=~ \frac{1}{2} \left( \partial_\mu \phi^{a_1 a_2} \right)^2 ~+~
\frac{g}{6}  f^{a_1 b_1 c_1} f^{a_2 b_2 c_2} \, \phi^{a_1 a_2} \phi^{b_1 b_2} 
\phi^{c_1 c_2} ~.
\label{lagbi}
\end{equation}
Given that the fields take values in the group $G_1$~$\times$~$G_2$ we use a
different notation from \cite{19,20} as we will carry out a more general 
analysis subsequently. Therefore we note that the numerical label on the 
adjoint indices will correspond to the label on the respective subgroups in the
overall symmetry group. The Roman letter that carries that label is the one
that is summed over in any repetition. Moreover these indices will run over a
set whose dimension is the dimension of the adjoint representation of the
respective group and denoted by $N_i$. So for instance 
$1$~$\leq$~$a_1$~$\leq$~$N_1$ and $1$~$\leq$~$b_2$~$\leq$~$N_2$ or in more 
general terms $1$~$\leq$~$a_i$,~$b_i$,~$c_i$,~$d_i$~$\leq$~$N_i$ for each $i$. 
In other words 
\begin{equation}
\delta^{a_i a_i} ~=~ N_{i} 
\label{groupN}
\end{equation}
where there is never a sum over the repeated label $i$ which indicates the 
specific group. If $G_1$~$=$~$SU(\Nc)$ for example then 
$N_1$~$=$~$\Nc^2$~$-$~$1$ corresponding to the dimension of the adjoint
representation. The respective structure functions of each group appearing in
(\ref{lagbi}) are $f^{a_i b_i c_i}$ for $i$~$=$~$1$ and $2$. As we will be
carrying out loop computations it is worth discussing related group theory 
quantities that will appear later. For example, we use a compact notation for
the Casimirs of each with
\begin{equation}
f^{a_i c_i d_i} f^{b_i c_i d_i} ~=~ C_{i} \delta^{a_i b_i} 
\label{groupC}
\end{equation}
for each $i$. Ordinarily one uses $C_A$ or $C_2(G)$ where $A$ denotes the
adjoint representation of the group $G$ for what we now denote by $C_i$. It is
not necessary however to include the representation designation since the 
adjoint will be used throughout the article unless stated otherwise. Beyond the
first few loop orders higher rank group Casimirs will appear. This was noted 
when the four loop $\beta$-function of Quantum Chromodynamics (QCD) was 
determined in \cite{22} and a comprehensive study was undertaken in \cite{23}
of general Lie group Casimirs in the context of perturbative computations. We 
briefly summarize the relevant aspects of that are needed here. For instance 
the fully symmetric rank $4$ tensor defined by, \cite{23}, 
\begin{equation}
d_{R_i}^{a_i b_i c_i d_i} ~=~ \frac{1}{6} \mbox{Tr} 
\left( T_{R_i}^{a_i} T_{R_i}^{(b_i} T_{R_i}^{c_i} T_{R_i}^{d_i)}
\right) 
\label{rank4}
\end{equation}
will arise. Here we revert momentarily to representation $R_i$ of the group
$G_i$ as (\ref{rank4}) involves the group generators $T^{a_i}_{R_i}$. Within 
the computation the contracted product of these tensors will produce additional
group Casmirs independent of $C_i$. At four loops in Yang-Mills theory the only
combination that appears turns out to be the simple product, \cite{23}, 
\begin{equation}
d_{(i) 44} ~=~ d_A^{a_i b_i c_i d_i} d_A^{a_i b_i c_i d_i} 
\label{groupd}
\end{equation}
in our notation where the bracketed label is used to avoid any potential 
ambiguity with the tensor rank when products of more groups are considered and 
$A$ denotes the adjoint representation. In the four loop QCD $\beta$-function, 
\cite{23}, by contrast, products of $d_{R_i}^{a_i b_i c_i d_i}$ in the 
fundamental and adjoint representations also arise. As a reference point for 
results that appear later we recall that for $SU(\Nc)$, \cite{23}, 
\begin{equation}
d_{(i) 44} ~=~ \frac{N_c^2[N_c^2+36]}{24} N_i
\end{equation}
and we have not substituted the explicit value for $N_i$ as that quantity 
appears in the results for a general Lie group which is what we use throughout.
Higher rank tensors beyond (\ref{rank4}) have been discussed in \cite{23}. A 
secondary motivation for studying the renormalization of (\ref{lagbi}) at large
loop order is to ascertain whether such rank $4$ tensor Casimirs first appear 
at the same loop order as that of QCD or not. In terms of other aspects of 
(\ref{lagbi}) for ease of comparison we retain the conventions that were used 
in \cite{24}. In particular in \cite{24} the sign of the coupling constant $g$ 
was opposite to that used in earlier work by others such as \cite{3,25,26,27}. 
There ought not to be difficulty in translating where necessary.

Before renormalizing (\ref{lagbi}) we recall our notation. First if we denote 
bare entities with a subscript ${}_{\mbox{\footnotesize{o}}}$ then their
relation to the renormalized counterparts are
\begin{equation}
\phi^{a_1 a_2}_{\mbox{\footnotesize{o}}} ~=~ 
\sqrt{Z_\phi} \, \phi^{a_1 a_2} ~~,~~ g_{\mbox{\footnotesize{o}}} ~=~ Z_g \, g
\label{rencon}
\end{equation}
in six dimensions. However we will dimensionally regularize (\ref{lagbi}) in
$d$~$=$~$6$~$-$~$2\epsilon$ dimensions and determine the renormalization group
functions in the $\MSbar$ scheme. To find $Z_\phi$ and $Z_g$ to four loop order
we have to compute the $2$- and $3$-point functions and we will follow the 
algorithm used in \cite{24} where more details of the technicalities of this 
exercise can be found. However given the presence of the structure constants in
the interaction we have had to adapt that method to determine $Z_g$ in 
particular. For example there are $540$ Feynman graphs to evaluate for the four
loop vertex function which is a large number to handle. To circumvent this a 
shortcut was exploited which was to generate these and lower loop vertex graphs
from the $2$-point graphs by applying a simple mapping to each propagator. In 
other words setting 
\begin{equation}
\frac{1}{k^2} ~\rightarrow~ \frac{1}{k^2} ~+~ \frac{\lambda}{(k^2)^2}
\label{propmap}
\end{equation}
and retaining only terms with zero or one power of the parameter $\lambda$,
means that the $O(\lambda)$ terms will formally correspond to $3$-point 
function graphs where one external momentum is set to zero, \cite{24}. In six 
dimensions this is infrared safe and evaluating all the graphs that are 
$O(\lambda)$ will produce the full $3$-point function to that loop order. While
this simple mapping is the essence of what one needs to do at the level of the 
graph generation it is not sufficient for (\ref{lagbi}) as the group theory 
factors need to be accommodated. To achieve this we adapt (\ref{propmap}) by 
including the group structure of the propagator as well as the inserted vertex 
to produce the mapping for (\ref{lagbi}) 
\begin{equation}
\frac{1}{k^2} \delta^{a_1 b_1} \delta^{a_2 b_2} ~\rightarrow~ 
\frac{1}{k^2} \delta^{a_1 b_1} \delta^{a_2 b_2} ~+~ 
\frac{\lambda g}{(k^2)^2} f^{a_1 b_1 c^e_1} f^{a_2 b_2 c^e_2} 
\label{propmapgen}
\end{equation}
where $c_i^e$ are the external indices of the inserted leg of the generated
$3$-point function. One benefit of using this technique is that there are only
$64$ four loop graphs in the $2$-point function and the insertion does not
change the underlying graph topology. This is important and leads to a more
efficient computation since the same integration subroutine for that topology
can be used to determine the divergences of the $2$-point graph and its
associated $3$-point one where there is a nullified insertion on each 
propagator. While we have not included a mass term in (\ref{lagbi}) we can
still compute the anomalous dimension of the mass by inserting the mass
operator $\half \phi^{a_1 a_2} \phi^{a_1 a_2}$ in a $2$-point function. The
renormalization constant associated with this operator is equivalent to that of
the mass itself which we denote by $Z_m$. Therefore we can directly use
(\ref{propmap}) to determine $Z_m$ as well as $Z_g$.

The computational strategy to evaluate the graphs of the $2$- and $3$-point 
functions is to use the Laporta integration by parts algorithm, \cite{28}.
This constructs relations between a set of Feynman integrals that can be
algebraically solved in such a way that all the integrals are related to a
relatively small set. These are termed the master integrals and have to be
evaluated directly. In our case since all the $2$-point four loop master
integrals are available in four dimensions, \cite{29}, it was possible to
connect these to the ones that emerge in our six dimensional computation,
\cite{24}. This is achieved by the Tarasov method, \cite{30,31}, whereby
integrals in $d$-dimensions can be related to the $(d+2)$-dimesional integral
with same topology and other topologies where one or more edges have been
removed. Therefore the four loop six dimensional masters were deduced in
\cite{24}. To effect the Laporta algorithm we used the {\sc Reduze}
implementation, \cite{32,33}. A useful feature of the package is that it allows
the database that is generated to be written in the symbolic language 
{\sc Form}, \cite{34,35}. This is important since we have written an automatic 
programme in {\sc Form} to carry out the full computation. In particular the 
contributing Feynman graphs are generated with the {\sc Qgraf} package,
\cite{36}, and the topology mapping appended. This allows the automatic 
programme to proceed since the integration of each topology follows a separate 
path. The final stage is the summation of all the graphs and the implementation
of the automatic renormalization to deduce $Z_\phi$, $Z_m$ and $Z_g$. To do 
this we follow \cite{37} which means that the graphs are evaluated with bare 
parameters with the renormalization constants (\ref{rencon}) being introduced 
at the end. For instance the $2$-point function is multiplied overall by 
$Z_\phi$ which allows one to deduce the unknown counterterms. Equally for the 
extraction of the mass and coupling constant renormalization constants the
parameter $\lambda$ in each of (\ref{propmap}) and (\ref{propmapgen}) are 
multiplied by $Z_m$ and $Z_g \sqrt{Z_\phi}$ respectively. One major tool that 
was used to carry out the manipulation of the large number of structure 
functions present at each vertex was the {\tt color.h} package written in 
{\sc Form} and available from \cite{34}. It encodes the group theory discussed 
in \cite{23} in an efficient way particularly for the three and four loop 
graphs. 

\sect{Results.}

Having outlined our computational strategy we can now record the outcome of
determining $Z_\phi$, $Z_m$ and $Z_g$ that lead to the respective
renormalization group functions $\gamma_\phi(g)$, $\gamma_m(g)$ and $\beta(g)$.
First the anomalous dimension of the scalar field is
\begin{eqnarray}
\gamma_\phi(g) &=& -~ \frac{g^2}{12} C_{1} C_{2} ~-~ 
\frac{5g^4}{432} C_{1}^2 C_{2}^2 ~-~ \frac{827g^6}{248832} C_{1}^3 C_{2}^3
\nonumber \\
&& +~ \left[ 1032 \zeta_3 C_{1}^4 C_{2}^4 N_{1} N_{2} 
+ 108 \zeta_4 C_{1}^4 C_{2}^4 N_{1} N_{2}
- 960 \zeta_5 C_{1}^4 C_{2}^4 N_{1} N_{2} - 943 C_{1}^4 C_{2}^4 N_{1} N_{2}
\right. \nonumber \\
&& \left. ~~~~
- 576 \zeta_3 C_{1}^4 d_{(2) 44} N_{1} - 2592 \zeta_4 C_{1}^4 d_{(2) 44} N_{1} 
- 11520 \zeta_5 C_{1}^4 d_{(2) 44} N_{1} 
\right. \nonumber \\
&& \left. ~~~~
+ 14976 C_{1}^4 d_{(2) 44} N_{1} - 576 \zeta_3 C_{2}^4 d_{(1) 44} N_{2} 
- 2592 \zeta_4 C_{2}^4 d_{(1) 44} N_{2} 
\right. \nonumber \\
&& \left. ~~~~
- 11520 \zeta_5 C_{2}^4 d_{(1) 44} N_{2} 
+ 14976 C_{2}^4 d_{(1) 44} N_{2}
+ 96768 \zeta_3 d_{(1) 44} d_{(2) 44} 
\right. \nonumber \\
&& \left. ~~~~
+ 62208 \zeta_4 d_{(1) 44} d_{(2) 44} 
- 138240 \zeta_5 d_{(1) 44} d_{(2) 44} \right]  
\frac{g^8}{497664 N_{1} N_{2}} ~+~ O(g^{10}) 
\label{bianom}
\end{eqnarray}
where $\zeta_z$ denotes the Riemann zeta function and we note that the rank $4$
group tensors first appear at four loops. For the mass anomalous dimension we
find 
\begin{eqnarray}
\gamma_m(g) &=& -~ \frac{g^2}{2} C_{1} C_{2} +
\frac{5g^4}{48} C_{1}^2 C_{2}^2 +
\left[ 432 \zeta_3 - 2203 \right] \frac{C_{1}^3 C_{2}^3 g^6}{13824} 
\nonumber \\
&& +~ \left[ 127764 \zeta_3 C_{1}^4 C_{2}^4 N_{1} N_{2} 
+ 1944 \zeta_4 C_{1}^4 C_{2}^4 N_{1} N_{2} 
- 251640 \zeta_5 C_{1}^4 C_{2}^4 N_{1} N_{2} 
\right. \nonumber \\
&& \left. ~~~~
+ 255517 C_{1}^4 C_{2}^4 N_{1} N_{2} + 88128 \zeta_3 C_{1}^4 d_{(2)44} N_{1} 
- 46656 \zeta_4 C_{1}^4 d_{(2)44} N_{1}
\right. \nonumber \\
&& \left. ~~~~
- 492480 \zeta_5 C_{1}^4 d_{(2)44} N_{1} + 409536 C_{1}^4 d_{(2)44} N_{1}
+ 88128 \zeta_3 C_{2}^4 d_{(1)44} N_{2} 
\right. \nonumber \\
&& \left. ~~~~
- 46656 \zeta_4 C_{2}^4 d_{(1)44} N_{2}
- 492480 \zeta_5 C_{2}^4 d_{(1)44} N_{2} + 409536 C_{2}^4 d_{(1)44} N_{2}
\right. \nonumber \\
&& \left. ~~~~
+ 3110400 \zeta_3 d_{(1)44} d_{(2)44} + 1119744 \zeta_4 d_{(1)44} d_{(2)44}
\right. \nonumber \\
&& \left. ~~~~
+ 622080 \zeta_5 d_{(1)44} d_{(2)44} \right] 
\frac{g^8}{1492992 N_{1} N_{2}} ~+~ 
O(g^{10})
\label{bimass}
\end{eqnarray} 
where the higher order Casimirs first appear at the same order. 

To complete the set the four loop $\beta$-function is 
\begin{eqnarray}
\beta(g) &=& -~ \frac{5g^5}{1152} C_{1}^2 C_{2}^2 \nonumber \\
&& +~ \left[ 108 \zeta_3 C_{1}^4 C_{2}^4 N_{1} N_{2} 
- 161 C_{1}^4 C_{2}^4 N_{1} N_{2} - 2592 \zeta_3 C_{1}^4 d_{(2) 44} N_{1} 
+ 2592 C_{1}^4 d_{(2) 44} N_{1}
\right. \nonumber \\
&& \left. ~~~~
- 2592 \zeta_3 C_{2}^4 d_{(1) 44} N_{2} + 2592 C_{2}^4 d_{(1) 44} N_{2}
+ 62208 \zeta_3 d_{(1) 44} d_{(2) 44} \right] 
\frac{g^7}{41472 C_{1} C_{2} N_{1} N_{2}}
\nonumber \\
&& +~ 
\left[ -~ 368928 \zeta_3 C_{1}^4 C_{2}^4 N_{1} N_{2} 
+ 518400 \zeta_5 C_{1}^4 C_{2}^4 N_{1} N_{2} 
- 101089 C_{1}^4 C_{2}^4 N_{1} N_{2} 
\right. \nonumber \\
&& \left. ~~~~
- 3483648 \zeta_3 C_{1}^4 d_{(2) 44} N_{1}
+ 6220800 \zeta_5 C_{1}^4 d_{(2) 44} N_{1} - 2363904 C_{1}^4 d_{(2) 44} N_{1}
\right. \nonumber \\
&& \left. ~~~~
- 3483648 \zeta_3 C_{2}^4 d_{(1) 44} N_{2} 
+ 6220800 \zeta_5 C_{2}^4 d_{(1) 44} N_{2}
- 2363904 C_{2}^4 d_{(1) 44} N_{2} 
\right. \nonumber \\
&& \left. ~~~~
- 95551488 \zeta_3 d_{(1) 44} d_{(2) 44} 
+ 119439360 \zeta_5 d_{(1) 44} d_{(2) 44} 
\right] \frac{g^9}{23887872 N_{1} N_{2}} \nonumber \\
&& +~ O(g^{11}) ~.
\label{bibeta}
\end{eqnarray}
This has the unusual feature in that the first non-zero term is at two loops 
rather than one loop. This is not the first or only case of the first term of a
renormalization group function in a fully renormalizable field theory being 
absent. For instance, while the field anomalous dimension in four dimensional 
$\phi^4$ theory is zero at one loop this is for the simple reason that the only
graph contributing to the $2$-point function is a snail. Therefore it is 
independent of the external momentum and its divergence contributes to the mass
renormalization only. Here the situation is different in that the only one loop 
graph of the $3$-point vertex is divergent but the residue of the simple pole 
is exactly cancelled by the contribution from the wave function 
renormalization. This is not the case for other symmetry decorations of the 
scalar field in scalar $\phi^3$ theory in six dimensions, \cite{3,26,27}. This 
curious property has an interesting consequence which is that since the 
coefficient of the now leading two loop term of $\beta(g)$ is negative then the
theory is asymptotically free. Ordinarily when this is a feature of other field
theories it is purely from the one loop term, \cite{1,2}. We note at this point
that this sign of the two loop term would have emerged irrespective of the 
coupling constant sign convention alluded to earlier. One comment that deserves
mention at this point concerns the scheme dependence of this particular 
$\beta$-function. Even though the one loop term is zero the three loop term of 
$\beta(g)$ still depends on the renormalization scheme. Unlike the other two 
renormalization group functions the rank $4$ Casimirs first appear at three 
loop in the $\beta$-function rather than four. This is one order earlier than 
that of QCD, \cite{22}.

To gain more insight into the consequences of their being no one loop term of
the $\beta$-function it is worth focussing on the case when both groups $G_1$
and $G_2$ are the same which we will denote by $G$. In this case 
(\ref{bianom}), (\ref{bimass}) and (\ref{bibeta}) become 
\begin{eqnarray}
\gamma_\phi^{G\times G}(g) &=&
-~ \frac{C_1^2 g^2}{12}  
- \frac{5 C_1^4 g^4}{432} 
- \frac{827 C_1^6 g^6}{248832} \nonumber \\
&& +~ \left[ 1032 \zeta_3 C_1^8 N_1^2 
- 1152 \zeta_3 C_1^4 d_{(1)44} N_1 
+ 96768 \zeta_3 d_{(1)44}^2 
+ 108 \zeta_4 C_1^8 N_1^2 
\right. \nonumber \\
&& \left. ~~~~
- 5184 \zeta_4 C_1^4 d_{(1)44} N_1 
+ 62208 \zeta_4 d_{(1)44}^2 
- 960 \zeta_5 C_1^8 N_1^2 
- 23040 \zeta_5 C_1^4 d_{(1)44} N_1 
\right. \nonumber \\
&& \left. ~~~~
- 138240 \zeta_5 d_{(1)44}^2 
- 943 C_1^8 N_1^2 
+ 29952 C_1^4 d_{(1)44} N_1 \right] \frac{g^8}{497664 N_1^2} \nonumber \\
&& +~ O(g^{10})
\end{eqnarray}
\begin{eqnarray}
\gamma_m^{G\times G}(g) &=&
-~ \frac{C_1^2 g^2}{2}  
+ \frac{5 C_1^4 g^4}{48}  
+ C_1^6 \left[ 432 \zeta_3 - 2203 \right] \frac{g^6}{13824} \nonumber \\
&& +~ \left[ 127764 \zeta_3 C_1^8 N_1^2 
+ 176256 \zeta_3 C_1^4 d_{(1)44} N_1 
+ 3110400 \zeta_3 d_{(1)44}^2 
+ 1944 \zeta_4 C_1^8 N_1^2 
\right. \nonumber \\
&& \left. ~~~~
- 93312 \zeta_4 C_1^4 d_{(1)44} N_1 
+ 1119744 \zeta_4 d_{(1)44}^2 
- 251640 \zeta_5 C_1^8 N_1^2 
\right. \nonumber \\
&& \left. ~~~~
- 984960 \zeta_5 C_1^4 d_{(1)44} N_1 
+ 622080 \zeta_5 d_{(1)44}^2 
+ 255517 C_1^8 N_1^2 
\right. \nonumber \\
&& \left. ~~~~
+ 819072 C_1^4 d_{(1)44} N_1 \right] \frac{g^8}{1492992 N_1^2} ~+~ O(g^{10}) 
\end{eqnarray} 
and
\begin{eqnarray}
\beta^{G\times G}(g) &=&
-~ \frac{5 C_1^4 g^5}{1152} \nonumber \\
&& +~ \left[ 108 \zeta_3 C_1^8 N_1^2 
- 5184 \zeta_3 C_1^4 d_{(1)44} N_1 
+ 62208 \zeta_3 d_{(1)44}^2 
- 161 C_1^8 N_1^2 
\right. \nonumber \\
&& \left. ~~~~
+ 5184 C_1^4 d_{(1)44} N_1 \right] \frac{g^7}{41472 C_1^2 N_1^2} \nonumber \\
&& +~ \left[ 
- 368928 \zeta_3 C_1^8 N_1^2 
- 6967296 \zeta_3 C_1^4 d_{(1)44} N_1 
- 95551488 \zeta_3 d_{(1)44}^2 
\right. \nonumber \\
&& \left. ~~~~
+ 518400 \zeta_5 C_1^8 N_1^2 
+ 12441600 \zeta_5 C_1^4 d_{(1)44} N_1 
+ 119439360 \zeta_5 d_{(1)44}^2 
\right. \nonumber \\
&& \left. ~~~~
- 101089 C_1^8 N_1^2 
- 4727808 C_1^4 d_{(1)44} N_1 \right] \frac{g^9}{23887872 N_1^2} ~+~
O(g^{11}) ~.
\label{betabiG}
\end{eqnarray}
Specifying to the group $SU(3)$ we deduce
\begin{eqnarray}
\gamma^{SU(3)\times SU(3)}_\phi(g) &=& -~ \frac{3}{42} g^2 ~-~ 
\frac{15}{16} g^4 ~-~ \frac{2481}{1024} g^6 \nonumber \\ 
&& +~ 27 \left[ 4992 \zeta_3 + 1728 \zeta_4 - 11760 \zeta_5 + 5297 \right]
\frac{g^8}{2048} ~+~ O(g^{10}) \nonumber \\
\gamma^{SU(3)\times SU(3)}_m(g) &=& -~ \frac{9}{2} g^2 ~+~ 
\frac{135}{16} g^4 ~+~ 27 \left[ 432 \zeta_3 - 2203 \right] \frac{g^6}{512} 
\nonumber \\ 
&& +~ 9 \left[ 299484 \zeta_3 + 31104 \zeta_4 - 429840 \zeta_5 + 426157 \right]
\frac{g^8}{2048} ~+~ O(g^{10}) \nonumber \\
\beta^{SU(3)\times SU(3)}(g) &=& -~ \frac{45}{128} g^5 ~+~ 
9 \left[ 1728 \zeta_3 + 919 \right] \frac{g^7}{512} \nonumber \\ 
&& +~ 9 \left[ 8294400 \zeta_5 - 5967648 \zeta_3 - 1086049 \right] 
\frac{g^9}{32768} ~+~ O(g^{11})
\end{eqnarray} 
or
\begin{eqnarray}
\gamma^{SU(3)\times SU(3)}_\phi(g) &=& -~ 0.750000 g^2 ~-~ 0.937500 g^4 ~-~ 
2.422852 g^6  ~+~ 12.836235 g^8 ~+~ O(g^{10}) \nonumber \\
\gamma^{SU(3)\times SU(3)}_m(g) &=& -~ 4.500000 g^2 ~+~ 8.437500 g^4 ~-~ 
88.789469 g^6  ~+~ 1644.017718 g^8 \nonumber \\
&& +~ O(g^{10}) \nonumber \\ 
\beta^{SU(3)\times SU(3)}(g) &=& -~ 0.351562 g^5 ~+~ 52.666775 g^7 ~+~ 
93.711209 g^9 ~+~ O(g^{11})
\label{rgesu3su3}
\end{eqnarray} 
numerically. From (\ref{rgesu3su3}) it is clear that there is a Banks-Zaks
fixed point, \cite{38} stemming from the opposite signs of the first two terms
of the $\beta$-function. In fact it is also the case that a similar fixed
point is present for $SU(N)$~$\times$~$SU(N)$. Strictly the fixed point of 
\cite{38} derives from the one and two loop terms but we use it in the sense of
the first two non-zero terms here. Moreover the value of the critical coupling 
for $N$~$=$~$3$ only changes by around $1\%$ when solving 
$\beta^{SU(3)\times SU(3)}(g)$~$=$~$0$ at three and four loops. One of the 
reasons for providing this example is to show another interesting consequence 
of the absence of the one loop term. While the renormalization group functions 
are scheme dependent one can derive renormalization group invariants from them.
These are critical exponents that are the evaluation of the functions at a 
non-trivial fixed point. One important such fixed point is the Wilson-Fisher 
one, \cite{39,40}, where the critical coupling is defined by setting the 
$d$-dimensional $\beta$-function to zero and denoted by $g_\ast$. So in 
$d$~$=$~$6$~$-$~$2\epsilon$ dimensions we have 
\begin{eqnarray}
\eta^{SU(3)\times SU(3)} &=& -~ \frac{2i\sqrt{5}}{5} \sqrt{\epsilon} ~+~
\frac{2}{25} \left[ 576 \zeta_3 + 323 \right] \epsilon \nonumber \\
&& +~ \sqrt{5} i \left[ 1327104 \zeta_3^2 + 779232 \zeta_3 + 921600 \zeta_5 
+ 271945
\right] \frac{\epsilon^{\threehalves}}{500} \nonumber \\
&& +~ \left[ 311040 \zeta_4 - 143327232 \zeta_3^2  - 96163200 \zeta_3 
- 85060800 \zeta_5 - 31005017 \right] \frac{\epsilon^2}{6750} \nonumber \\
&& +~ O(\epsilon^{\fivehalves}) \nonumber \\
\eta_m^{SU(3)\times SU(3)} &=& -~ \frac{12i\sqrt{5}}{5} \sqrt{\epsilon} ~+~
16 \left[ 432 \zeta_3 + 211 \right] \frac{\epsilon}{25} \nonumber \\
&& +~ \frac{9i\sqrt{5}}{250} \left[ 442368 \zeta_3^2 + 233664 \zeta_3 
+ 307200 \zeta_5 + 79175 \right] \epsilon^{\threehalves} \nonumber \\
&& +~ 2 \left[ 110854656 \zeta_3^2 + 55734372 \zeta_3 + 155520 \zeta_4 
+ 60058800 \zeta_5 + 15279107 \right] \frac{\epsilon^2}{1125} \nonumber \\
&& +~ O(\epsilon^{\fivehalves}) \nonumber \\
\omega^{SU(3)\times SU(3)} &=& 2 \epsilon ~-~ \frac{2i\sqrt{5}}{5}
\left[ 1728 \zeta_3 + 919 \right] \epsilon^{\threehalves} \nonumber \\
&& +~ \left[ 35831808 \zeta_3^2 + 8274528 \zeta_3 + 41472000 \zeta_5 + 4704487
\right] \frac{\epsilon^2}{2250} ~+~ O(\epsilon^{\fivehalves})
\end{eqnarray} 
where $\eta$~$=$~$\gamma_\phi(g_\ast)$, $\eta_m$~$=$~$\gamma_m(g_\ast)$ and
$\omega$~$=$~$2\beta^\prime(g_\ast)$. The main key difference between these
exponents and those from models where there is a non-zero one loop term is that
the expansion is a function of $\sqrt{\epsilon}$ rather than $\epsilon$. In
addition the exponents are complex but this is due to having assumed $\epsilon$
is real and positive. If $\epsilon$ were real and negative then the exponents 
are real above six dimensions. 

Finally we close this section by recalling that solving the $\beta$-function as
a differential equation determines the functional dependence of the running 
coupling constant with the renormalization scale $\mu$. Therefore we can 
formally compare the running coupling constants in the conventional case where 
asymptotic freedom is determined by the one loop $\beta$-function with that for
(\ref{lagbi}). For instance, if we formally define two $\beta$-functions by
\begin{eqnarray}
\beta_1(g_1) &=& -~ \beta_1 g_1^3 ~+~ O(g_1^5) \nonumber \\ 
\beta_2(g_2) &=& -~ \beta_2 g_2^5 ~+~ O(g_2^7) 
\end{eqnarray}
then we have 
\begin{equation}
g_1^2(\mu) ~=~ -~ \frac{1}{\beta_1 \ln \left( \mu^2/\Lambda_1^2 \right)}
\end{equation}
for the more conventional one loop $\beta$-function. By contrast solving the
second case we find
\begin{equation}
g_2^2(\mu) ~=~ -~ 
\frac{1}{\sqrt{2\beta_2 \ln \left( \mu^2/\Lambda_2^2 \right)}}
\end{equation}
where $\Lambda_1$ and $\Lambda_2$ are the constants of integration. Clearly 
both running coupling constants have the same general behaviour in that they 
tend to zero as $\mu$~$\to$~$\infty$. However in the latter case where the one 
loop $\beta$-function term is absent, the coupling constant tends to zero at a 
much slower rate. So if this model, or one with the same property, was realized
in Nature the constituent particles would only be effectively free at 
significantly high energies. 

\sect{Quartic adjoint.}

The absence of a one loop term in the $\beta$-function of (\ref{lagbi}) is an
interesting property. In order to see whether this property is common to more 
general scalar $\phi^3$ theories with adjoint decorations we have repeated the 
renormalization exercise for (\ref{lagbi}) for what we will term the quartic 
adjoint theory with Lagrangian 
\begin{equation}
L ~=~ \frac{1}{2} \left( \partial_\mu \phi^{a_1 a_2 a_3 a_4} \right)^2 ~+~
\frac{g}{6}  f^{a_1 b_1 c_1} f^{a_2 b_2 c_2} f^{a_3 b_3 c_3} f^{a_4 b_4 c_4} 
\phi^{a_1 a_2 a_3 a_4} \phi^{b_1 b_2 b_3 b_4} \phi^{c_1 c_2 c_3 c_4} ~.
\label{lagquar}
\end{equation}
Here we have a scalar field which takes values in the group
$G_1$~$\times$~$G_2$~$\times$~$G_3$~$\times$~$G_4$ and in particular the 
interaction involves the adjoint representation of the group generators. In
(\ref{lagquar}) we use a similar notation to that introduced for (\ref{lagbi})
where there are now two additional labels due to the extra groups $G_3$ and
$G_4$. Equally the definition of the group Casimirs have an obvious natural 
extension of those given in (\ref{groupN}), (\ref{groupC}) and (\ref{groupd}).
We have followed the same process of renormalizing (\ref{lagquar}) as that for
(\ref{lagbi}) together with similar consistency checks. Therefore we move to
the discussion of the outcome. With the additional group structure it 
transpires that the four loop expressions for each of the renormalization group
functions are more involved than those of (\ref{lagbi}). These have been 
recorded in the Appendix. Instead we illustrate the structure in the simpler 
case of the group $G$~$\times$~$G$~$\times$~$G$~$\times$~$G$~$\equiv$~$G^4$ and
find
\begin{eqnarray}
\gamma_\phi^{G^4}(g) &=&
-~ \frac{C_1^4 g^2}{12}  
- \frac{19 C_1^8 g^4}{864}  
- \frac{40421 C_1^{12} g^6}{3981312} \nonumber \\
&& +~ \left[ 1910 \zeta_3 C_1^{16} N_1^4 
+ 2112 \zeta_3 C_1^{12} d_{(1)44} N_1^3 
+ 89856 \zeta_3 C_1^8 d_{(1)44}^2 N_1^2 
\right. \nonumber \\
&& \left. ~~~~
- 110592 \zeta_3 C_1^4 d_{(1)44}^3 N_1 
+ 4644864 \zeta_3 d_{(1)44}^4 
+ 9 \zeta_4 C_1^{16} N_1^4 
\right. \nonumber \\
&& \left. ~~~~
- 864 \zeta_4 C_1^{12} d_{(1)44} N_1^3 
+ 31104 \zeta_4 C_1^8 d_{(1)44}^2 N_1^2 
- 497664 \zeta_4 C_1^4 d_{(1)44}^3 N_1 
\right. \nonumber \\
&& \left. ~~~~
+ 2985984 \zeta_4 d_{(1)44}^4 
- 320 \zeta_5 C_1^{16} N_1^4 
- 15360 \zeta_5 C_1^{12} d_{(1)44} N_1^3 
\right. \nonumber \\
&& \left. ~~~~
- 276480 \zeta_5 C_1^8 d_{(1)44}^2 N_1^2 
- 2211840 \zeta_5 C_1^4 d_{(1)44}^3 N_1 
- 6635520 \zeta_5 d_{(1)44}^4 
\right. \nonumber \\
&& \left. ~~~~
- 150934 C_1^{16} N_1^4 
+ 14976 C_1^{12} d_{(1)44} N_1^3 
+ 179712 C_1^8 d_{(1)44}^2 N_1^2 
\right. \nonumber \\
&& \left. ~~~~
+ 2875392 C_1^4 d_{(1)44}^3 N_1 \right]
\frac{g^8}{23887872 N_1^4} ~+~ O(g^{10}) 
\end{eqnarray}
and
\begin{eqnarray}
\gamma_m^{G^4}(g) &=&
-~ \frac{C_1^4 g^2}{2}  
+ \frac{C_1^8 g^2}{96}  
+ \left[ 3024 \zeta_3 - 43537 \right] \frac{C_1^{12}g^6}{221184} \nonumber \\
&& +~ \left[  
- 619077 \zeta_3 C_1^{16} N_1^4 
+ 58176 \zeta_3 C_1^{12} d_{(1)44} N_1^3 
+ 1389312 \zeta_3 C_1^8 d_{(1)44}^2 N_1^2 
\right. \nonumber \\
&& \left. ~~~~
+ 5640192 \zeta_3 C_1^4 d_{(1)44}^3 N_1 
+ 49766400 \zeta_3 d_{(1)44}^4 
- 91800 \zeta_4 C_1^{16} N_1^4 
\right. \nonumber \\
&& \left. ~~~~
- 5184 \zeta_4 C_1^{12} d_{(1)44} N_1^3 
+ 186624 \zeta_4 C_1^8 d_{(1)44}^2 N_1^2 
- 2985984 \zeta_4 C_1^4 d_{(1)44}^3 N_1 
\right. \nonumber \\
&& \left. ~~~~
+ 17915904 \zeta_4 d_{(1)44}^4 
- 177630 \zeta_5 C_1^{16} N_1^4 
- 158400 \zeta_5 C_1^{12} d_{(1)44} N_1^3 
\right. \nonumber \\
&& \left. ~~~~
- 1762560 \zeta_5 C_1^8 d_{(1)44}^2 N_1^2 
- 31518720 \zeta_5 C_1^4 d_{(1)44}^3 N_1 
+ 9953280 \zeta_5 d_{(1)44}^4 
\right. \nonumber \\
&& \left. ~~~~
+ 2070250 C_1^{16} N_1^4 
+ 136512 C_1^{12} d_{(1)44} N_1^3 
+ 1638144 C_1^8 d_{(1)44}^2 N_1^2 
\right. \nonumber \\
&& \left. ~~~~
+ 26210304 C_1^4 d_{(1)44}^3 N_1 \right] 
\frac{g^8}{23887872 N_1^4} ~+~ O(g^{10}) 
\end{eqnarray}
for the field and mass anomalous dimensions. For the $\beta$-function we 
arrived at 
\begin{eqnarray}
\beta^{G^4}(g) &=&
-~ \frac{3 C_1^4 g^3}{32} 
- \frac{467 C_1^8g^5}{18432} \nonumber \\
&& +~ \left[ 48 \zeta_3 C_1^{16} N_1^4 
- 4608 \zeta_3 C_1^{12} d_{(1)44} N_1^3 
+ 165888 \zeta_3 C_1^8 d_{(1)44}^2 N_1^2 
\right. \nonumber \\
&& \left. ~~~~
- 2654208 \zeta_3 C_1^4 d_{(1)44}^3 N_1 
+ 15925248 \zeta_3 d_{(1)44}^4 
- 125981 C_1^{16} N_1^4 
\right. \nonumber \\
&& \left. ~~~~
+ 13824 C_1^{12} d_{(1)44} N_1^3 
+ 165888 C_1^8 d_{(1)44}^2 N_1^2 
\right. \nonumber \\
&& \left. ~~~~
+ 2654208 C_1^4 d_{(1)44}^3 N_1 \right] \frac{g^7}{10616832 C_1^4 N_1^4}
\nonumber \\
&& +~ \left[ 394304 \zeta_3 C_1^{16} N_1^4 
- 21494784 \zeta_3 C_1^{12} d_{(1)44} N_1^3 
- 224169984 \zeta_3 C_1^8 d_{(1)44}^2 N_1^2 
\right. \nonumber \\
&& \left. ~~~~
- 7582187520 \zeta_3 C_1^4 d_{(1)44}^3 N_1 
- 3301834752 \zeta_3 d_{(1)44}^4 
- 15552 \zeta_4 C_1^{16} N_1^4 
\right. \nonumber \\
&& \left. ~~~~
+ 1492992 \zeta_4 C_1^{12} d_{(1)44} N_1^3 
- 53747712 \zeta_4 C_1^8 d_{(1)44}^2 N_1^2 
+ 859963392 \zeta_4 C_1^4 d_{(1)44}^3 N_1 
\right. \nonumber \\
&& \left. ~~~~
- 5159780352 \zeta_4 d_{(1)44}^4 
+ 325120 \zeta_5 C_1^{16} N_1^4 
+ 18370560 \zeta_5 C_1^{12} d_{(1)44} N_1^3 
\right. \nonumber \\
&& \left. ~~~~
+ 479969280 \zeta_5 C_1^8 d_{(1)44}^2 N_1^2 
+ 4636016640 \zeta_5 C_1^4 d_{(1)44}^3 N_1 
+ 24418713600 \zeta_5 d_{(1)44}^4 
\right. \nonumber \\
&& \left. ~~~~
- 134800515 C_1^{16} N_1^4 
+ 15123456 C_1^{12} d_{(1)44} N_1^3 
+ 189444096 C_1^8 d_{(1)44}^2 N_1^2 
\right. \nonumber \\
&& \left. ~~~~
+ 2521497600 C_1^4 d_{(1)44}^3 N_1 \right]
\frac{g^9}{18345885696 N_1^4} ~+~ O(g^{11}) 
\label{betaquarG4}
\end{eqnarray}
and there is no Banks-Zaks fixed point. By contrast to (\ref{bibeta}) and the 
parallel simplification of (\ref{betabiG}) we note that there is a non-zero one
loop $\beta$-function coefficient unlike the bi-adjoint model. In the general 
group case this coefficient is a simple product of $C_i$ for $i$~$=$~$1$ to 
$4$.

In order to compare with the bi-adjoint case we note that specifying to the
group $SU(3)$ gives 
\begin{eqnarray}
\gamma^{SU(3)^4}_\phi(g) &=& -~ 6.750000 g^2 ~-~ 144.281250 g^4 ~-~ 
5395.552185 g^6  ~+~ 2.4888387 \times 10^5 g^8 \nonumber \\
&& +~ O(g^{10}) \nonumber \\
\gamma^{SU(3)^4}_m(g) &=& -~ 40.50000 g^2 ~+~ 68.343750 g^4 ~-~ 
95872.884627 g^6  ~+~ 2.842688 \times 10^6 g^8 \nonumber \\
&& +~ O(g^{10}) \nonumber \\ 
\beta^{SU(3)^4}(g) &=& -~ 7.593750 g^3 ~-~ 166.231934 g^5 ~-~ 
458.390411 g^7 ~-~ 95378.353885 g^9 \nonumber \\
&& +~ O(g^{11}) ~.
\end{eqnarray} 
With these we can illustrate the difference in the corresponding critical
exponents at the Wilson-Fisher fixed point in $d$~$=$~$6$~$-$~$2\epsilon$
dimensions by noting that
\begin{eqnarray}
\eta^{SU(3)^4} &=& \frac{4}{9} \epsilon + \frac{11}{729} \epsilon^2 
+ 2 [4608 \zeta_3 + 12295] \frac{\epsilon^3}{59049} \nonumber \\
&& +~ [ 88631400 \zeta_3 + 4478976 \zeta_4 - 115795200 \zeta_5 
+ 33923953 ] \frac{\epsilon^4}{19131876} ~+~ O(\epsilon^5)
\nonumber \\
\eta_m^{SU(3)^4} &=& \frac{8}{3} \epsilon + \frac{1006}{243} \epsilon^2 
+ [ - 22392 \zeta_3 + 684799 ] \frac{\epsilon^3}{19683} \nonumber \\
&& +~ [33249552 \zeta_3 - 10882512 \zeta_4 - 288886560 \zeta_5
+ 1436683139 ] \frac{\epsilon^4}{6377292} ~+~ O(\epsilon^5)
\nonumber \\
\omega^{SU(3)^4} &=& \epsilon - \frac{467}{324} \epsilon^2 
+ [ - 18432 \zeta_3 + 28807 ] \frac{\epsilon^3}{26244} \nonumber \\
&& +~ [ - 467532144 \zeta_3 - 17915904 \zeta_4 + 508069440 \zeta_5 
- 192428981 ] \frac{\epsilon^4}{17006112} \nonumber \\
&& +~ O(\epsilon^5) ~.
\end{eqnarray} 
The non-zero one loop $\beta$-function coefficient produces the standard 
$\epsilon$ expansion in contrast to the bi-adjoint case where the exponents 
depend on $\sqrt{\epsilon}$.

Having considered a second scalar theory with a group theory structure similar
to that of (\ref{lagbi}) which does not have a zero one loop $\beta$-function
coefficient it is worth trying to understand how this arises for (\ref{lagbi}). 
There are two parts to determining the renormalization constant $Z_g$ that
leads to the $\beta$-function. These are the divergences from the $2$- and 
$3$-point functions. The former produces the value for $Z_\phi$ directly 
whereas the divergences of the latter do not immediately give $Z_g$. Instead it
gives the combination $Z_g Z_\phi^{\frac{3}{2}}$. So for $Z_g$ to have no 
simple pole at one loop means that the divergence from the $3$-point function 
must exactly match that of $Z_\phi$ multiplied by $\frac{3}{2}$. From the 
explicit computation we find that the residue of the one loop simple pole of 
the $2$-point function is $-~\frac{1}{12} C_1 C_2$ whereas that for the 
$3$-point function is $\frac{1}{8} C_1 C_2$. These are clearly in the required 
ratio. By contrast the respective numbers for (\ref{lagquar}) are 
$-~\frac{1}{12} C_1 C_2 C_3 C_4$ and $\frac{1}{32} C_1 C_2 C_3 C_4$. Combining 
these to deduce $Z_g$ at one loop gives the correct coefficient of 
$-~\frac{3}{32} C_1 C_2 C_3 C_4$ of the general $\beta$-function. Aside from 
the additional group theory factors the only discrepancy between both models is
in the coefficient of the divergence from the $3$-point function which is 
different by a factor of $\frac{1}{4}$. This is the origin of why 
(\ref{bibeta}) has no one loop term and rests in the group theory deriving from
the one loop triangle graph which is the sole contribution at this loop order. 
For each subgroup of the symmetry group the group theory contribution is
\begin{equation}
f^{a_i p_i q_i} f^{b_i r_i q_i} f^{c_i p_i r_i} ~=~ 
\frac{1}{2} C_i f^{a_i b_i c_i} ~. 
\end{equation}
So this relation, derived from the Jacobi identity in the adjoint
representation, gives a factor of $\frac{1}{2}$ for each group $G_i$ to the 
residue of the simple pole of the $3$-point function. As there is one factor of
$\frac{1}{2}$ from the actual integration over the loop momentum then for the 
most general group $G_1$~$\times$~$\ldots$~$\times$~$G_n$ the $3$-point 
function simple pole has a residue of $\frac{1}{2^{n+1}} \prod_{i=1}^n C_i$. 
Hence for this general group the one loop coefficient of the $\beta$-function, 
denoted by $\beta_1(n)$, will be 
\begin{equation}
\beta_1(n) ~=~ \left[ \frac{1}{2^{n+1}} ~-~ \frac{1}{8} \right] 
\prod_{i=1}^n C_i  
\label{beta1}
\end{equation}
which is a monotonically decreasing function and defined for all integers
$n$~$\neq$~$1$. The exception is because one has a free field theory for 
$n$~$=$~$1$ since the interaction is $f^{a_1 b_1 c_1} \phi^{a_1} \phi^{b_1} 
\phi^{c_1}$ and vanishes due to the antisymmetry of the structure constants. 
Clearly $\beta_1(2)$~$=$~$0$ and so the curiosity of (\ref{lagbi}) being 
asymptotically free as a consequence of the two loop term of the 
$\beta$-function is purely due to a group theory property. We note that the 
value of $\beta_1(0)$ is consistent with the known low order $\beta$-function 
of the pure $\phi^3$ theory, \cite{3,25}. Equally $\beta_1(4)$ is consistent 
with (\ref{betaquarG4}) and (\ref{betaquar}).

\sect{Discussion.}

Scalar $\phi^3$ theory has played an important role as a toy model in quantum 
field theory for many decades. For instance any Feynman graph generated from 
the basic cubic interaction can in turn generate the basic topologies that
can occur in higher $n$-point interactions. This is achieved by formally
deleting propagators in the graph theory sense and hence represents the initial
point for combinatoric studies in quantum field theory. Where the theory has 
limitations in physics applications is that its critical dimension is six 
rather than four. However as noted earlier certain properties of scalar 
$\phi^3$ theory are similar to the more involved field theories in four
dimensions and hence the six dimensional model can be used to explore ideas. In
this article we have studied a interesting modification whereby the scalar 
field is in a bi-adjoint representation of Lie groups. This is motivated by the
double copy relation between Yang-Mills and on-shell gravity. While the studies
of \cite{18,19,20,21} examined classical solutions to the scalar theory it has 
turned out that the six dimensional field theory has a peculiar property. It is
unusual that asymptotic freedom is a consequence of the {\em two} loop term of 
the $\beta$-function rather than the first. However that is the case for 
(\ref{lagbi}). In studying the consequences it appears to be unique in the 
class of extensions that would be termed multi-adjoint as the analysis we 
carried out for the quartic adjoint demonstrates. It is not clear whether there
is a parallel theory in four dimensions that is asymptotically free due to the 
two loop $\beta$-function for which the bi-adjoint six dimensional scalar field
theory is the underlying laboratory. It was noted in \cite{41} that a necessary
condition for this is non-abelian gauge fields.

\vspace{1cm}
\noindent
{\bf Acknowledgements.} This work was supported in part by a DFG Mercator 
Fellowship.

\appendix

\sect{Full results for quartic adjoint.}

In this appendix we record the full expressions for the renormalization group
functions of the quartic adjoint scalar theory which uses similar notation to
that used for the parallel expressions of the bi-adjoint case. First the 
field anomalous and mass anomalous dimensions are 
\begin{eqnarray}
\gamma_\phi(g) &=&
-~ C_1 C_2 C_3 C_4 \frac{g^2}{12}
- 19 C_1^2 C_2^2 C_3^2 C_4^2 \frac{g^4}{864} 
- 40421 C_1^3 C_2^3 C_3^3 C_4^3 \frac{g^6}{3981312} \nonumber \\
&& +~ \left[ 1910 \zeta_3 C_1^4 C_2^4 C_3^4 C_4^4 N_1 N_2 N_3 N_4 
+ 9 \zeta_4 C_1^4 C_2^4 C_3^4 C_4^4 N_1 N_2 N_3 N_4 
\right. \nonumber \\
&& \left. ~~~~
- 320 \zeta_5 C_1^4 C_2^4 C_3^4 C_4^4 N_1 N_2 N_3 N_4 
- 150934 C_1^4 C_2^4 C_3^4 C_4^4 N_1 N_2 N_3 N_4 
\right. \nonumber \\
&& \left. ~~~~
+ 528 \zeta_3 C_1^4 C_2^4 C_3^4 d_{(4)44} N_1 N_2 N_3 
- 216 \zeta_4 C_1^4 C_2^4 C_3^4 d_{(4)44} N_1 N_2 N_3 
\right. \nonumber \\
&& \left. ~~~~
- 3840 \zeta_5 C_1^4 C_2^4 C_3^4 d_{(4)44} N_1 N_2 N_3 
+ 3744 C_1^4 C_2^4 C_3^4 d_{(4)44} N_1 N_2 N_3 
\right. \nonumber \\
&& \left. ~~~~
+ 528 \zeta_3 C_1^4 C_2^4 C_4^4 d_{(3)44} N_1 N_2 N_4 
- 216 \zeta_4 C_1^4 C_2^4 C_4^4 d_{(3)44} N_1 N_2 N_4 
\right. \nonumber \\
&& \left. ~~~~
- 3840 \zeta_5 C_1^4 C_2^4 C_4^4 d_{(3)44} N_1 N_2 N_4 
+ 3744 C_1^4 C_2^4 C_4^4 d_{(3)44} N_1 N_2 N_4 
\right. \nonumber \\
&& \left. ~~~~
+ 14976 \zeta_3 C_1^4 C_2^4 d_{(3)44} d_{(4)44} N_1 N_2 
+ 5184 \zeta_4 C_1^4 C_2^4 d_{(3)44} d_{(4)44} N_1 N_2 
\right. \nonumber \\
&& \left. ~~~~
- 46080 \zeta_5 C_1^4 C_2^4 d_{(3)44} d_{(4)44} N_1 N_2 
+ 29952 C_1^4 C_2^4 d_{(3)44} d_{(4)44} N_1 N_2 
\right. \nonumber \\
&& \left. ~~~~
+ 528 \zeta_3 C_1^4 C_3^4 C_4^4 d_{(2)44} N_1 N_3 N_4 
- 216 \zeta_4 C_1^4 C_3^4 C_4^4 d_{(2)44} N_1 N_3 N_4 
\right. \nonumber \\
&& \left. ~~~~
- 3840 \zeta_5 C_1^4 C_3^4 C_4^4 d_{(2)44} N_1 N_3 N_4 
+ 3744 C_1^4 C_3^4 C_4^4 d_{(2)44} N_1 N_3 N_4 
\right. \nonumber \\
&& \left. ~~~~
+ 14976 \zeta_3 C_1^4 C_3^4 d_{(2)44} d_{(4)44} N_1 N_3 
+ 5184 \zeta_4 C_1^4 C_3^4 d_{(2)44} d_{(4)44} N_1 N_3 
\right. \nonumber \\
&& \left. ~~~~
- 46080 \zeta_5 C_1^4 C_3^4 d_{(2)44} d_{(4)44} N_1 N_3 
+ 29952 C_1^4 C_3^4 d_{(2)44} d_{(4)44} N_1 N_3 
\right. \nonumber \\
&& \left. ~~~~
+ 14976 \zeta_3 C_1^4 C_4^4 d_{(2)44} d_{(3)44} N_1 N_4 
+ 5184 \zeta_4 C_1^4 C_4^4 d_{(2)44} d_{(3)44} N_1 N_4 
\right. \nonumber \\
&& \left. ~~~~
- 46080 \zeta_5 C_1^4 C_4^4 d_{(2)44} d_{(3)44} N_1 N_4 
+ 29952 C_1^4 C_4^4 d_{(2)44} d_{(3)44} N_1 N_4 
\right. \nonumber \\
&& \left. ~~~~
- 27648 \zeta_3 C_1^4 d_{(2)44} d_{(3)44} d_{(4)44} N_1 
- 124416 \zeta_4 C_1^4 d_{(2)44} d_{(3)44} d_{(4)44} N_1 
\right. \nonumber \\
&& \left. ~~~~
- 552960 \zeta_5 C_1^4 d_{(2)44} d_{(3)44} d_{(4)44} N_1 
+ 718848 C_1^4 d_{(2)44} d_{(3)44} d_{(4)44} N_1 
\right. \nonumber \\
&& \left. ~~~~
+ 528 \zeta_3 C_2^4 C_3^4 C_4^4 d_{(1)44} N_2 N_3 N_4 
- 216 \zeta_4 C_2^4 C_3^4 C_4^4 d_{(1)44} N_2 N_3 N_4 
\right. \nonumber \\
&& \left. ~~~~
- 3840 \zeta_5 C_2^4 C_3^4 C_4^4 d_{(1)44} N_2 N_3 N_4 
+ 3744 C_2^4 C_3^4 C_4^4 d_{(1)44} N_2 N_3 N_4 
\right. \nonumber \\
&& \left. ~~~~
+ 14976 \zeta_3 C_2^4 C_3^4 d_{(1)44} d_{(4)44} N_2 N_3 
+ 5184 \zeta_4 C_2^4 C_3^4 d_{(1)44} d_{(4)44} N_2 N_3 
\right. \nonumber \\
&& \left. ~~~~
- 46080 \zeta_5 C_2^4 C_3^4 d_{(1)44} d_{(4)44} N_2 N_3 
+ 29952 C_2^4 C_3^4 d_{(1)44} d_{(4)44} N_2 N_3 
\right. \nonumber \\
&& \left. ~~~~
+ 14976 \zeta_3 C_2^4 C_4^4 d_{(1)44} d_{(3)44} N_2 N_4 
+ 5184 \zeta_4 C_2^4 C_4^4 d_{(1)44} d_{(3)44} N_2 N_4 
\right. \nonumber \\
&& \left. ~~~~
- 46080 \zeta_5 C_2^4 C_4^4 d_{(1)44} d_{(3)44} N_2 N_4 
+ 29952 C_2^4 C_4^4 d_{(1)44} d_{(3)44} N_2 N_4 
\right. \nonumber \\
&& \left. ~~~~
- 27648 \zeta_3 C_2^4 d_{(1)44} d_{(3)44} d_{(4)44} N_2 
- 124416 \zeta_4 C_2^4 d_{(1)44} d_{(3)44} d_{(4)44} N_2 
\right. \nonumber \\
&& \left. ~~~~
- 552960 \zeta_5 C_2^4 d_{(1)44} d_{(3)44} d_{(4)44} N_2 
+ 718848 C_2^4 d_{(1)44} d_{(3)44} d_{(4)44} N_2 
\right. \nonumber \\
&& \left. ~~~~
+ 14976 \zeta_3 C_3^4 C_4^4 d_{(1)44} d_{(2)44} N_3 N_4 
+ 5184 \zeta_4 C_3^4 C_4^4 d_{(1)44} d_{(2)44} N_3 N_4 
\right. \nonumber \\
&& \left. ~~~~
- 46080 \zeta_5 C_3^4 C_4^4 d_{(1)44} d_{(2)44} N_3 N_4 
+ 29952 C_3^4 C_4^4 d_{(1)44} d_{(2)44} N_3 N_4 
\right. \nonumber \\
&& \left. ~~~~
- 27648 \zeta_3 C_3^4 d_{(1)44} d_{(2)44} d_{(4)44} N_3 
- 124416 \zeta_4 C_3^4 d_{(1)44} d_{(2)44} d_{(4)44} N_3 
\right. \nonumber \\
&& \left. ~~~~
- 552960 \zeta_5 C_3^4 d_{(1)44} d_{(2)44} d_{(4)44} N_3 
+ 718848 C_3^4 d_{(1)44} d_{(2)44} d_{(4)44} N_3 
\right. \nonumber \\
&& \left. ~~~~
- 27648 \zeta_3 C_4^4 d_{(1)44} d_{(2)44} d_{(3)44} N_4 
- 124416 \zeta_4 C_4^4 d_{(1)44} d_{(2)44} d_{(3)44} N_4 
\right. \nonumber \\
&& \left. ~~~~
- 552960 \zeta_5 C_4^4 d_{(1)44} d_{(2)44} d_{(3)44} N_4 
+ 718848 C_4^4 d_{(1)44} d_{(2)44} d_{(3)44} N_4 
\right. \nonumber \\
&& \left. ~~~~
+ 4644864 \zeta_3 d_{(1)44} d_{(2)44} d_{(3)44} d_{(4)44} 
+ 2985984 \zeta_4 d_{(1)44} d_{(2)44} d_{(3)44} d_{(4)44} 
\right. \nonumber \\
&& \left. ~~~~
- 6635520 \zeta_5 d_{(1)44} d_{(2)44} d_{(3)44} d_{(4)44} \right]
\frac{g^8}{23887872 N_1 N_2 N_3 N_4} ~+~ O(g^{10}) 
\end{eqnarray}
and
\begin{eqnarray}
\gamma_m(g) &=& 
-~ C_1 C_2 C_3 C_4 \frac{g^2}{2} 
+ C_1^2 C_2^2 C_3^2 C_4^2 \frac{g^2}{96}  
+ C_1^3 C_2^3 C_3^3 C_4^3 \left[ 3024 \zeta_3 - 43537 \right] 
\frac{g^6}{221184} \nonumber \\
&& +~ \left[ 
- 619077 \zeta_3 C_1^4 C_2^4 C_3^4 C_4^4 N_1 N_2 N_3 N_4 
- 91800 \zeta_4 C_1^4 C_2^4 C_3^4 C_4^4 N_1 N_2 N_3 N_4 
\right. \nonumber \\
&& \left. ~~~~
- 177630 \zeta_5 C_1^4 C_2^4 C_3^4 C_4^4 N_1 N_2 N_3 N_4 
+ 2070250 C_1^4 C_2^4 C_3^4 C_4^4 N_1 N_2 N_3 N_4 
\right. \nonumber \\
&& \left. ~~~~
+ 14544 \zeta_3 C_1^4 C_2^4 C_3^4 d_{(4)44} N_1 N_2 N_3 
- 1296 \zeta_4 C_1^4 C_2^4 C_3^4 d_{(4)44} N_1 N_2 N_3 
\right. \nonumber \\
&& \left. ~~~~
- 39600 \zeta_5 C_1^4 C_2^4 C_3^4 d_{(4)44} N_1 N_2 N_3 
+ 34128 C_1^4 C_2^4 C_3^4 d_{(4)44} N_1 N_2 N_3 
\right. \nonumber \\
&& \left. ~~~~
+ 14544 \zeta_3 C_1^4 C_2^4 C_4^4 d_{(3)44} N_1 N_2 N_4 
- 1296 \zeta_4 C_1^4 C_2^4 C_4^4 d_{(3)44} N_1 N_2 N_4 
\right. \nonumber \\
&& \left. ~~~~
- 39600 \zeta_5 C_1^4 C_2^4 C_4^4 d_{(3)44} N_1 N_2 N_4 
+ 34128 C_1^4 C_2^4 C_4^4 d_{(3)44} N_1 N_2 N_4 
\right. \nonumber \\
&& \left. ~~~~
+ 231552 \zeta_3 C_1^4 C_2^4 d_{(3)44} d_{(4)44} N_1 N_2 
+ 31104 \zeta_4 C_1^4 C_2^4 d_{(3)44} d_{(4)44} N_1 N_2 
\right. \nonumber \\
&& \left. ~~~~
- 293760 \zeta_5 C_1^4 C_2^4 d_{(3)44} d_{(4)44} N_1 N_2 
+ 273024 C_1^4 C_2^4 d_{(3)44} d_{(4)44} N_1 N_2 
\right. \nonumber \\
&& \left. ~~~~
+ 14544 \zeta_3 C_1^4 C_3^4 C_4^4 d_{(2)44} N_1 N_3 N_4 
- 1296 \zeta_4 C_1^4 C_3^4 C_4^4 d_{(2)44} N_1 N_3 N_4 
\right. \nonumber \\
&& \left. ~~~~
- 39600 \zeta_5 C_1^4 C_3^4 C_4^4 d_{(2)44} N_1 N_3 N_4 
+ 34128 C_1^4 C_3^4 C_4^4 d_{(2)44} N_1 N_3 N_4 
\right. \nonumber \\
&& \left. ~~~~
+ 231552 \zeta_3 C_1^4 C_3^4 d_{(2)44} d_{(4)44} N_1 N_3 
+ 31104 \zeta_4 C_1^4 C_3^4 d_{(2)44} d_{(4)44} N_1 N_3 
\right. \nonumber \\
&& \left. ~~~~
- 293760 \zeta_5 C_1^4 C_3^4 d_{(2)44} d_{(4)44} N_1 N_3 
+ 273024 C_1^4 C_3^4 d_{(2)44} d_{(4)44} N_1 N_3 
\right. \nonumber \\
&& \left. ~~~~
+ 231552 \zeta_3 C_1^4 C_4^4 d_{(2)44} d_{(3)44} N_1 N_4 
+ 31104 \zeta_4 C_1^4 C_4^4 d_{(2)44} d_{(3)44} N_1 N_4 
\right. \nonumber \\
&& \left. ~~~~
- 293760 \zeta_5 C_1^4 C_4^4 d_{(2)44} d_{(3)44} N_1 N_4 
+ 273024 C_1^4 C_4^4 d_{(2)44} d_{(3)44} N_1 N_4 
\right. \nonumber \\
&& \left. ~~~~
+ 1410048 \zeta_3 C_1^4 d_{(2)44} d_{(3)44} d_{(4)44} N_1 
- 746496 \zeta_4 C_1^4 d_{(2)44} d_{(3)44} d_{(4)44} N_1 
\right. \nonumber \\
&& \left. ~~~~
- 7879680 \zeta_5 C_1^4 d_{(2)44} d_{(3)44} d_{(4)44} N_1 
+ 6552576 C_1^4 d_{(2)44} d_{(3)44} d_{(4)44} N_1 
\right. \nonumber \\
&& \left. ~~~~
+ 14544 \zeta_3 C_2^4 C_3^4 C_4^4 d_{(1)44} N_2 N_3 N_4 
- 1296 \zeta_4 C_2^4 C_3^4 C_4^4 d_{(1)44} N_2 N_3 N_4 
\right. \nonumber \\
&& \left. ~~~~
- 39600 \zeta_5 C_2^4 C_3^4 C_4^4 d_{(1)44} N_2 N_3 N_4 
+ 34128 C_2^4 C_3^4 C_4^4 d_{(1)44} N_2 N_3 N_4 
\right. \nonumber \\
&& \left. ~~~~
+ 231552 \zeta_3 C_2^4 C_3^4 d_{(1)44} d_{(4)44} N_2 N_3 
+ 31104 \zeta_4 C_2^4 C_3^4 d_{(1)44} d_{(4)44} N_2 N_3 
\right. \nonumber \\
&& \left. ~~~~
- 293760 \zeta_5 C_2^4 C_3^4 d_{(1)44} d_{(4)44} N_2 N_3 
+ 273024 C_2^4 C_3^4 d_{(1)44} d_{(4)44} N_2 N_3 
\right. \nonumber \\
&& \left. ~~~~
+ 231552 \zeta_3 C_2^4 C_4^4 d_{(1)44} d_{(3)44} N_2 N_4 
+ 31104 \zeta_4 C_2^4 C_4^4 d_{(1)44} d_{(3)44} N_2 N_4 
\right. \nonumber \\
&& \left. ~~~~
- 293760 \zeta_5 C_2^4 C_4^4 d_{(1)44} d_{(3)44} N_2 N_4 
+ 273024 C_2^4 C_4^4 d_{(1)44} d_{(3)44} N_2 N_4 
\right. \nonumber \\
&& \left. ~~~~
+ 1410048 \zeta_3 C_2^4 d_{(1)44} d_{(3)44} d_{(4)44} N_2 
- 746496 \zeta_4 C_2^4 d_{(1)44} d_{(3)44} d_{(4)44} N_2 
\right. \nonumber \\
&& \left. ~~~~
- 7879680 \zeta_5 C_2^4 d_{(1)44} d_{(3)44} d_{(4)44} N_2 
+ 6552576 C_2^4 d_{(1)44} d_{(3)44} d_{(4)44} N_2 
\right. \nonumber \\
&& \left. ~~~~
+ 231552 \zeta_3 C_3^4 C_4^4 d_{(1)44} d_{(2)44} N_3 N_4 
+ 31104 \zeta_4 C_3^4 C_4^4 d_{(1)44} d_{(2)44} N_3 N_4 
\right. \nonumber \\
&& \left. ~~~~
- 293760 \zeta_5 C_3^4 C_4^4 d_{(1)44} d_{(2)44} N_3 N_4 
+ 273024 C_3^4 C_4^4 d_{(1)44} d_{(2)44} N_3 N_4 
\right. \nonumber \\
&& \left. ~~~~
+ 1410048 \zeta_3 C_3^4 d_{(1)44} d_{(2)44} d_{(4)44} N_3 
- 746496 \zeta_4 C_3^4 d_{(1)44} d_{(2)44} d_{(4)44} N_3 
\right. \nonumber \\
&& \left. ~~~~
- 7879680 \zeta_5 C_3^4 d_{(1)44} d_{(2)44} d_{(4)44} N_3 
+ 6552576 C_3^4 d_{(1)44} d_{(2)44} d_{(4)44} N_3 
\right. \nonumber \\
&& \left. ~~~~
+ 1410048 \zeta_3 C_4^4 d_{(1)44} d_{(2)44} d_{(3)44} N_4 
- 746496 \zeta_4 C_4^4 d_{(1)44} d_{(2)44} d_{(3)44} N_4 
\right. \nonumber \\
&& \left. ~~~~
- 7879680 \zeta_5 C_4^4 d_{(1)44} d_{(2)44} d_{(3)44} N_4 
+ 6552576 C_4^4 d_{(1)44} d_{(2)44} d_{(3)44} N_4 
\right. \nonumber \\
&& \left. ~~~~
+ 49766400 \zeta_3 d_{(1)44} d_{(2)44} d_{(3)44} d_{(4)44} 
+ 17915904 \zeta_4 d_{(1)44} d_{(2)44} d_{(3)44} d_{(4)44} 
\right. \nonumber \\
&& \left. ~~~~
+ 9953280 \zeta_5 d_{(1)44} d_{(2)44} d_{(3)44} d_{(4)44} \right]
\frac{g^8}{23887872 N_1 N_2 N_3 N_4} ~+~ O(g^{10}) ~.
\end{eqnarray}
Finally, the $\beta$-function is
\begin{eqnarray}
\beta(g) &=& -~ 3 C_1 C_2 C_3 C_4 \frac{g^3}{32}  
- 467 C_1^2 C_2^2 C_3^2 C_4^2 \frac{g^5}{18432} \nonumber \\
&& +~ \left[ 48 \zeta_3 C_1^4 C_2^4 C_3^4 C_4^4 N_1 N_2 N_3 N_4 
- 125981 C_1^4 C_2^4 C_3^4 C_4^4 N_1 N_2 N_3 N_4 
\right. \nonumber \\
&& \left. ~~~~
- 1152 \zeta_3 C_1^4 C_2^4 C_3^4 d_{(4)44} N_1 N_2 N_3 
+ 3456 C_1^4 C_2^4 C_3^4 d_{(4)44} N_1 N_2 N_3 
\right. \nonumber \\
&& \left. ~~~~
- 1152 \zeta_3 C_1^4 C_2^4 C_4^4 d_{(3)44} N_1 N_2 N_4 
+ 3456 C_1^4 C_2^4 C_4^4 d_{(3)44} N_1 N_2 N_4 
\right. \nonumber \\
&& \left. ~~~~
+ 27648 \zeta_3 C_1^4 C_2^4 d_{(3)44} d_{(4)44} N_1 N_2 
+ 27648 C_1^4 C_2^4 d_{(3)44} d_{(4)44} N_1 N_2 
\right. \nonumber \\
&& \left. ~~~~
- 1152 \zeta_3 C_1^4 C_3^4 C_4^4 d_{(2)44} N_1 N_3 N_4 
+ 3456 C_1^4 C_3^4 C_4^4 d_{(2)44} N_1 N_3 N_4 
\right. \nonumber \\
&& \left. ~~~~
+ 27648 \zeta_3 C_1^4 C_3^4 d_{(2)44} d_{(4)44} N_1 N_3 
+ 27648 C_1^4 C_3^4 d_{(2)44} d_{(4)44} N_1 N_3 
\right. \nonumber \\
&& \left. ~~~~
+ 27648 \zeta_3 C_1^4 C_4^4 d_{(2)44} d_{(3)44} N_1 N_4 
+ 27648 C_1^4 C_4^4 d_{(2)44} d_{(3)44} N_1 N_4 
\right. \nonumber \\
&& \left. ~~~~
- 663552 \zeta_3 C_1^4 d_{(2)44} d_{(3)44} d_{(4)44} N_1 
+ 663552 C_1^4 d_{(2)44} d_{(3)44} d_{(4)44} N_1 
\right. \nonumber \\
&& \left. ~~~~
- 1152 \zeta_3 C_2^4 C_3^4 C_4^4 d_{(1)44} N_2 N_3 N_4 
+ 3456 C_2^4 C_3^4 C_4^4 d_{(1)44} N_2 N_3 N_4 
\right. \nonumber \\
&& \left. ~~~~
+ 27648 \zeta_3 C_2^4 C_3^4 d_{(1)44} d_{(4)44} N_2 N_3 
+ 27648 C_2^4 C_3^4 d_{(1)44} d_{(4)44} N_2 N_3 
\right. \nonumber \\
&& \left. ~~~~
+ 27648 \zeta_3 C_2^4 C_4^4 d_{(1)44} d_{(3)44} N_2 N_4 
+ 27648 C_2^4 C_4^4 d_{(1)44} d_{(3)44} N_2 N_4 
\right. \nonumber \\
&& \left. ~~~~
- 663552 \zeta_3 C_2^4 d_{(1)44} d_{(3)44} d_{(4)44} N_2 
+ 663552 C_2^4 d_{(1)44} d_{(3)44} d_{(4)44} N_2 
\right. \nonumber \\
&& \left. ~~~~
+ 27648 \zeta_3 C_3^4 C_4^4 d_{(1)44} d_{(2)44} N_3 N_4 
+ 27648 C_3^4 C_4^4 d_{(1)44} d_{(2)44} N_3 N_4 
\right. \nonumber \\
&& \left. ~~~~
- 663552 \zeta_3 C_3^4 d_{(1)44} d_{(2)44} d_{(4)44} N_3 
+ 663552 C_3^4 d_{(1)44} d_{(2)44} d_{(4)44} N_3 
\right. \nonumber \\
&& \left. ~~~~
- 663552 \zeta_3 C_4^4 d_{(1)44} d_{(2)44} d_{(3)44} N_4 
+ 663552 C_4^4 d_{(1)44} d_{(2)44} d_{(3)44} N_4 
\right. \nonumber \\
&& \left. ~~~~
+ 15925248 \zeta_3 d_{(1)44} d_{(2)44} d_{(3)44} d_{(4)44} \right]
\frac{g^7}{10616832 C_1 C_2 C_3 C_4 N_1 N_2 N_3 N_4} \nonumber \\
&& +~ \left[ 394304 \zeta_3 C_1^4 C_2^4 C_3^4 C_4^4 N_1 N_2 N_3 N_4 
- 15552 C_1^4 C_2^4 C_3^4 C_4^4 N_1 N_2 N_3 N_4 \zeta_4 
\right. \nonumber \\
&& \left. ~~~~
+ 325120 C_1^4 C_2^4 C_3^4 C_4^4 N_1 N_2 N_3 N_4 \zeta_5 
- 134800515 C_1^4 C_2^4 C_3^4 C_4^4 N_1 N_2 N_3 N_4 
\right. \nonumber \\
&& \left. ~~~~
- 5373696 \zeta_3 C_1^4 C_2^4 C_3^4 d_{(4)44} N_1 N_2 N_3 
+ 373248 \zeta_4 C_1^4 C_2^4 C_3^4 d_{(4)44} N_1 N_2 N_3 
\right. \nonumber \\
&& \left. ~~~~
+ 4592640 \zeta_5 C_1^4 C_2^4 C_3^4 d_{(4)44} N_1 N_2 N_3 
+ 3780864 C_1^4 C_2^4 C_3^4 d_{(4)44} N_1 N_2 N_3 
\right. \nonumber \\
&& \left. ~~~~
- 5373696 \zeta_3 C_1^4 C_2^4 C_4^4 d_{(3)44} N_1 N_2 N_4 
+ 373248 \zeta_4 C_1^4 C_2^4 C_4^4 d_{(3)44} N_1 N_2 N_4 
\right. \nonumber \\
&& \left. ~~~~
+ 4592640 \zeta_5 C_1^4 C_2^4 C_4^4 d_{(3)44} N_1 N_2 N_4 
+ 3780864 C_1^4 C_2^4 C_4^4 d_{(3)44} N_1 N_2 N_4 
\right. \nonumber \\
&& \left. ~~~~
- 37361664 \zeta_3 C_1^4 C_2^4 d_{(3)44} d_{(4)44} N_1 N_2 
- 8957952 \zeta_4 C_1^4 C_2^4 d_{(3)44} d_{(4)44} N_1 N_2 
\right. \nonumber \\
&& \left. ~~~~
+ 79994880 \zeta_5 C_1^4 C_2^4 d_{(3)44} d_{(4)44} N_1 N_2 
+ 31574016 C_1^4 C_2^4 d_{(3)44} d_{(4)44} N_1 N_2 
\right. \nonumber \\
&& \left. ~~~~
- 5373696 \zeta_3 C_1^4 C_3^4 C_4^4 d_{(2)44} N_1 N_3 N_4 
+ 373248 \zeta_4 C_1^4 C_3^4 C_4^4 d_{(2)44} N_1 N_3 N_4 
\right. \nonumber \\
&& \left. ~~~~
+ 4592640 \zeta_5 C_1^4 C_3^4 C_4^4 d_{(2)44} N_1 N_3 N_4 
+ 3780864 C_1^4 C_3^4 C_4^4 d_{(2)44} N_1 N_3 N_4 
\right. \nonumber \\
&& \left. ~~~~
- 37361664 \zeta_3 C_1^4 C_3^4 d_{(2)44} d_{(4)44} N_1 N_3 
- 8957952 \zeta_4 C_1^4 C_3^4 d_{(2)44} d_{(4)44} N_1 N_3 
\right. \nonumber \\
&& \left. ~~~~
+ 79994880 \zeta_5 C_1^4 C_3^4 d_{(2)44} d_{(4)44} N_1 N_3 
+ 31574016 C_1^4 C_3^4 d_{(2)44} d_{(4)44} N_1 N_3 
\right. \nonumber \\
&& \left. ~~~~
- 37361664 \zeta_3 C_1^4 C_4^4 d_{(2)44} d_{(3)44} N_1 N_4 
- 8957952 \zeta_4 C_1^4 C_4^4 d_{(2)44} d_{(3)44} N_1 N_4 
\right. \nonumber \\
&& \left. ~~~~
+ 79994880 \zeta_5 C_1^4 C_4^4 d_{(2)44} d_{(3)44} N_1 N_4 
+ 31574016 C_1^4 C_4^4 d_{(2)44} d_{(3)44} N_1 N_4 
\right. \nonumber \\
&& \left. ~~~~
- 1895546880 \zeta_3 C_1^4 d_{(2)44} d_{(3)44} d_{(4)44} N_1 
+ 214990848 \zeta_4 C_1^4 d_{(2)44} d_{(3)44} d_{(4)44} N_1 
\right. \nonumber \\
&& \left. ~~~~
+ 1159004160 \zeta_5 C_1^4 d_{(2)44} d_{(3)44} d_{(4)44} N_1 
+ 630374400 C_1^4 d_{(2)44} d_{(3)44} d_{(4)44} N_1 
\right. \nonumber \\
&& \left. ~~~~
- 5373696 \zeta_3 C_2^4 C_3^4 C_4^4 d_{(1)44} N_2 N_3 N_4 
+ 373248 \zeta_4 C_2^4 C_3^4 C_4^4 d_{(1)44} N_2 N_3 N_4 
\right. \nonumber \\
&& \left. ~~~~
+ 4592640 \zeta_5 C_2^4 C_3^4 C_4^4 d_{(1)44} N_2 N_3 N_4 
+ 3780864 C_2^4 C_3^4 C_4^4 d_{(1)44} N_2 N_3 N_4 
\right. \nonumber \\
&& \left. ~~~~
- 37361664 \zeta_3 C_2^4 C_3^4 d_{(1)44} d_{(4)44} N_2 N_3 
- 8957952 \zeta_4 C_2^4 C_3^4 d_{(1)44} d_{(4)44} N_2 N_3 
\right. \nonumber \\
&& \left. ~~~~
+ 79994880 \zeta_5 C_2^4 C_3^4 d_{(1)44} d_{(4)44} N_2 N_3 
+ 31574016 C_2^4 C_3^4 d_{(1)44} d_{(4)44} N_2 N_3 
\right. \nonumber \\
&& \left. ~~~~
- 37361664 \zeta_3 C_2^4 C_4^4 d_{(1)44} d_{(3)44} N_2 N_4 
- 8957952 \zeta_4 C_2^4 C_4^4 d_{(1)44} d_{(3)44} N_2 N_4 
\right. \nonumber \\
&& \left. ~~~~
+ 79994880 \zeta_5 C_2^4 C_4^4 d_{(1)44} d_{(3)44} N_2 N_4 
+ 31574016 C_2^4 C_4^4 d_{(1)44} d_{(3)44} N_2 N_4 
\right. \nonumber \\
&& \left. ~~~~
- 1895546880 \zeta_3 C_2^4 d_{(1)44} d_{(3)44} d_{(4)44} N_2 
+ 214990848 \zeta_4 C_2^4 d_{(1)44} d_{(3)44} d_{(4)44} N_2 
\right. \nonumber \\
&& \left. ~~~~
+ 1159004160 \zeta_5 C_2^4 d_{(1)44} d_{(3)44} d_{(4)44} N_2 
+ 630374400 C_2^4 d_{(1)44} d_{(3)44} d_{(4)44} N_2 
\right. \nonumber \\
&& \left. ~~~~
- 37361664 \zeta_3 C_3^4 C_4^4 d_{(1)44} d_{(2)44} N_3 N_4 
- 8957952 \zeta_4 C_3^4 C_4^4 d_{(1)44} d_{(2)44} N_3 N_4 
\right. \nonumber \\
&& \left. ~~~~
+ 79994880 \zeta_5 C_3^4 C_4^4 d_{(1)44} d_{(2)44} N_3 N_4 
+ 31574016 C_3^4 C_4^4 d_{(1)44} d_{(2)44} N_3 N_4 
\right. \nonumber \\
&& \left. ~~~~
- 1895546880 \zeta_3 C_3^4 d_{(1)44} d_{(2)44} d_{(4)44} N_3 
+ 214990848 \zeta_4 C_3^4 d_{(1)44} d_{(2)44} d_{(4)44} N_3 
\right. \nonumber \\
&& \left. ~~~~
+ 1159004160 \zeta_5 C_3^4 d_{(1)44} d_{(2)44} d_{(4)44} N_3 
+ 630374400 C_3^4 d_{(1)44} d_{(2)44} d_{(4)44} N_3 
\right. \nonumber \\
&& \left. ~~~~
- 1895546880 \zeta_3 C_4^4 d_{(1)44} d_{(2)44} d_{(3)44} N_4 
+ 214990848 \zeta_4 C_4^4 d_{(1)44} d_{(2)44} d_{(3)44} N_4 
\right. \nonumber \\
&& \left. ~~~~
+ 1159004160 \zeta_5 C_4^4 d_{(1)44} d_{(2)44} d_{(3)44} N_4 
+ 630374400 C_4^4 d_{(1)44} d_{(2)44} d_{(3)44} N_4 
\right. \nonumber \\
&& \left. ~~~~
- 3301834752 \zeta_3 d_{(1)44} d_{(2)44} d_{(3)44} d_{(4)44} 
- 5159780352 \zeta_4 d_{(1)44} d_{(2)44} d_{(3)44} d_{(4)44} 
\right. \nonumber \\
&& \left. ~~~~
+ 24418713600 \zeta_5 d_{(1)44} d_{(2)44} d_{(3)44} d_{(4)44} \right] 
\frac{g^9}{18345885696 N_1 N_2 N_3 N_4} ~. \nonumber \\
&& +~ O(g^{11}) 
\label{betaquar}
\end{eqnarray}

\end{document}